\documentclass[aps,prd,preprint,showpacs,preprintnumbers,nofootinbib]{revtex4}
\usepackage{epsfig}
\usepackage{latexsym,amsmath,amssymb,amstext}
\usepackage{float}
\usepackage{graphicx,xcolor}
\usepackage{soul}
\usepackage{slashed}
\usepackage[paperwidth=210mm,paperheight=297mm,centering,hmargin=2cm,vmargin=2.5cm]{geometry}

\newcommand{\be}{\begin{equation}}
\newcommand{\ee}{\end{equation}}
\newcommand{\bea}{\begin{eqnarray}}
\newcommand{\eea}{\end{eqnarray}}

\newcommand\bef{\begin{figure}}
\newcommand\eef[1]{\label{fg:#1}\end{figure}}
\newcommand\beq{\begin{equation}}
\newcommand\eeq[1]{\label{#1}\end{equation}}
\newcommand\beqa{\begin{eqnarray}}
\newcommand\eeqa[1]{\label{#1}\end{eqnarray}}
\newcommand\bet{\begin{table}}
\newcommand\eet[1]{\label{tb:#1}\end{table}}

\newcommand\fgn[1]{Figure \ref{fg:#1}}
\newcommand\eqn[1]{Eq.\ (\ref{#1})}
\newcommand\scn[1]{Section \ref{sec:#1}}

\begin{document}
\setstcolor{red}
\title{A curious behavior of three-dimensional lattice Dirac operators coupled to monopole background}
\author{Nikhil\ \surname{Karthik}}
\email{nkarthik@bnl.gov}
\affiliation{Physics Department, Brookhaven National Laboratory, Upton, New York, 11973-5000}
\author{Rajamani\ \surname{Narayanan}}
\email{rajamani.narayanan@fiu.edu}
\affiliation{Department of Physics, Florida International University, Miami, FL 33199.}

\begin{abstract}
We investigate numerically the effect of regulating fermions
in the presence of singular background fields in three dimensions.
For this, we couple free lattice fermions to a background compact U(1) gauge field
consisting of a monopole-anti-monopole pair of magnetic charge $\pm Q$
separated by a distance $s$ in a periodic $L^3$ lattice, and study the
low-lying eigenvalues of different lattice Dirac operators under a
continuum limit defined by taking $L\to\infty$ at fixed $s/L$.  As
the background gauge field is parity even, we look for a two-fold
degeneracy of the Dirac spectrum that is expected of a continuum-like
Dirac operator.  The naive-Dirac operator exhibits such a
parity-doubling, but breaks the degeneracy of the fermion-doubler
modes for the $Q$ lowest eigenvalues in the continuum limit.  The
Wilson-Dirac operator lifts the fermion-doublers but breaks the
parity-doubling in the $Q$ lowest modes even in the continuum limit.
The overlap-Dirac operator shows parity-doubling of all the modes
even at finite $L$ that is devoid
 of fermion-doubling,
and singles out as a properly regulated continuum Dirac operator
in the presence of singular gauge field configurations albeit with
a peculiar algorithmic issue.

\end{abstract}

\date{\today}
\maketitle

\section{Introduction}

Lattice regularization of non-compact QED~\cite{Hands:1989mv} in three dimensions is
defined by a non-compact action for the gauge fields, $\theta_\mu({\bf
n})\in {\mathbf R}$, on the link connecting ${\bf n}$ and ${\bf
n}+\hat\mu$  and the lattice fermions couple to $U(1)$ valued link
variables, $U_\mu({\bf n})=e^{i\theta_\mu({\bf n})}$. Monopoles are
suppressed in the continuum limit in such a regularization.  Recent
numerical analysis of non-compact QED in three dimensions with even
number of massless two component fermions shows that these theories
are scale invariant independent of the number of
flavors~\cite{Karthik:2015sgq,Karthik:2016ppr,Karthik:2017hol}.
It is natural to follow-up such a study with an analysis of compact
QED$_3$ where the lattice gauge action is the compact gauge
action~\cite{Armour:2011zx}.  When we attempted to numerically study this
theory using overlap-Dirac fermions, we found it be numerically formidable due to anomalously small
eigenvalues of the massive Wilson-Dirac kernel that is at the core of the
overlap-Dirac operator --- to contrast, for a smooth field, one would find 
the spectrum of a massive Wilson-Dirac operator to be gapped at least by the Wilson mass. This prompted us to consider the 
question as to what happens when the conventional lattice regulated fermions, 
which lead to universal results in the continuum limit over generic smooth gauge fields,
are coupled to a singular gauge field from a monopole; do operations at the level of 
lattice spacing, such as point-splitting used regularly in lattice regularization, have any effect in the presence of a 
Dirac string which is also one lattice spacing thick? We present related numerical observations in this paper.

Briefly, we recount some aspects of lattice fermions in three dimensions. 
The naive fermion operator $\slashed{D}$ obtained by using the discrete derivative operator is 
the simplest. As is well known, it leads to $2^d$(8 in three dimensions) fermions flavors. It is a well-motivated expectation
that there is flavor degeneracy in the continuum limit. There is a trivial two-fold degeneracy for naive-Dirac fermions~\cite{Kogut:1974ag,Gattringer:2010zz}
om the lattice and one copy is the staggered-Dirac fermion which is expected to realize a four fermion flavor theory in three dimensions.
If there is a four-fold degeneracy in the continuum limit,
one could possibly define a theory with the square root of the staggered-Dirac operator to study a two flavor parity invariant theory. Some continuum based reasoning provides
arguments as to why gauge field backgrounds with non-trivial topology might obstruct a well-defined continuum limit of a lattice theory with the fourth root
of the staggered-Dirac operator in even dimensions~\cite{Creutz:2008nk,Creutz:2008kb,Creutz:2008hx,Creutz:2007rk}.
It is possible monopole backgrounds in three dimensions suffer from similar effects.
The Wilson-Dirac operator is obtained by adding the Wilson-term $B$, which is 
irrelevant by naive power-counting, to the naive-operator $\slashed{D}$. That is, the massive
Wilson-Dirac operator is given by
\beq
X=-m_w+B + \slashed{D},
\eeq{wilsonx}
which lifts the mass of the seven of the doublers leaving only one physical fermion of lattice mass $m_w$
on smooth gauge fields.
The lattice fermion which is capable of reproducing the 
continuum symmetries, such as the U$(N)$ flavor symmetry in three-dimensional $N$-flavor QED$_3$, is the 
overlap-Dirac operator.
The central quantity that appears in the
overlap formalism~\cite{Neuberger:1997fp,Karthik:2016ppr} is the unitary operator $V$ defined as
\be
V = X \frac{1}{\sqrt{X^\dagger X}},
\ee
with the Wilson mass
$0 < m_w <2$ and the massless overlap 
operator is given by 
\be
D_o = \frac{1+V}{2};\qquad
D_o^\dagger D_o = \frac{ 2+V+V^\dagger}{4}\label{overlapddag}.
\ee
The instance where the otherwise irrelevant operators used in lattice 
regularization play significant roles is the parity anomaly~\cite{Redlich:1983dv,Niemi:1983rq,Coste:1989wf,Karthik:2015sza}.
Parity takes the naive-Dirac operator
$\slashed{D}$ to $\slashed{D}^\dagger = -\slashed{D}$;
the Wilson-Dirac operator $X$ transforms to
 $X^\dagger$ and the unitary operator $V$ to $V^\dagger$.
The phase of ${\det X}$ for $m_w=0$ is non-vanishing even in the continuum limit, even 
though the unregulated continuum massless Dirac operator is anti-hermitian.
This effect propagates itself to the non-vanishing phase of  $\det(1+V)$ of the massless 
overlap fermion.  Notwithstanding such effects in three-dimensions, we expect $X$ to
commute with $X^\dagger$ in the continuum limit, unless the gauge
field background is not smooth even in the continuum limit. Independent of
the nature of the gauge field background, $V$ and $V^\dagger$
commute. This places the overlap-Dirac operator closer to the
continuum Dirac operator compared to the Wilson-Dirac operator.
The domain-wall-Dirac operator formalism in three dimensions~\cite{Hands:2015qha,Hands:2015dyp} is expected to behave like the overlap-Dirac operator.

Having explained the lattice formalism, we return back to the
problem that motivated us to study the problem to be presented in
this paper. Following the conventions of~\cite{Karthik:2016ppr},
we will assume that $m_w>0$ in the region of interest and this will
lead us to the unconventional notation for Wilson-Dirac fermions,
namely; $m_w < 0$ will correspond to fermions with positive mass.
Since the operator $X^\dagger X$ can be viewed as the one for two
flavors of two component fermions that preserves parity, the sign
of the mass should not matter in the conventional approach to the
continuum limit. But, our attempts to study compact QED with
overlap-Dirac fermions failed due to several eigenvalues of $X^\dagger
X$ becoming very small for all values of $m_w \in (0,2)$. Furthermore,
we found the number of such anomalously small eigenvalues to grow
with the size of the three dimensional torus.

The above failure prompted us to study the low lying spectrum
of the following positive definite operators constructed out of
lattice operators; $\slashed{D}^\dagger
\slashed{D}$ for the  naive-Dirac operator; $X^\dagger X$ as a
function of $m_w$ for the Wilson-Dirac operator; and of the
$(1+V)(1+V^\dagger)$ for the overlap-Dirac operator in a controlled
background before proceeding to address an alternative approach to
the study of compact QED. As we will argue, the eigenvalues of such a positive definite 
operator is doubly degenerate in the continuum  in a monopole-anti-monopole
background, and hence serve as a promising observable to
look for any deviation of regulated lattice operator from the continuum one.
It is not possible to write down a
background gauge field that has a single monopole in a periodic
lattice but it is possible to write down one that has a
monopole-anti-monopole pair separated by a fixed distance. Such a
background was considered in a study of the monopole scaling
dimension~\cite{Karthik:2018rcg}. We will use a similar background
with a minor change to better fit it in a periodic lattice.

\section{The lattice monopole-anti-monopole field}\label{sec:mam}

A way to include the monopole-anti-monopole background field on the lattice is to integrate the 
continuum field ${\cal A}$ of a Dirac monopole-anti-monopole pair~\cite{Shnir:2005xx} over links 
joining site ${\bf x}$ to ${\bf x}+a$, where $a$ is the lattice spacing. That is, define a link variable
\be
\overline{\theta}_\mu(x)=\int_{\bf x}^{{\bf x}+a\hat\mu} dx'_\mu {\cal A}_\mu(\mathbf{x}'),
\ee
as given in~\cite{Karthik:2018rcg}.
The drawback of this approach is that periodicity of lattice forces artificial jumps 
in the gauge field across the ``boundaries". So we consider a better construction of the field 
on periodic lattice below.
\subsection{Monopole-anti-monopole field on periodic lattice}\label{sec:maml}
We implement the background gauge field that contains a
monopole-anti-monopole pair of integer charge $\pm Q$ and separated
by a length $s$ on a periodic lattice of length $L$
as defined by the following non-compact field strength $B_{\mu\nu}({\bf
n})$ at the lattice site ${\bf n}=(n_1,n_2,n_3)$:
\be
B_{23}({\bf n})=B_{31}({\bf n})=0;\qquad B_{12}({\bf n})=\begin{cases} 2\pi Q & n_1=n_2=\frac{L}{2};\quad 1\le n_3 \le s\cr 0 & {\rm otherwise},
\end{cases};\qquad {\bf n} \in [1,L]\label{diracs}.
\ee
That is, $B_{\mu\nu}({\bf n})$ denotes the non-compact field
strength on the directed plaquette defined by the corners ${\bf
n}$, ${\bf n}+\hat\mu$, ${\bf n}+\hat\mu+\hat\nu$ and ${\bf n}+\hat\nu$
traversed in the anti-clockwise direction.  As constructed, the
monopole charge density is
\be
Q({\bf n}) = \frac{1}{4\pi} \sum_{\mu\nu\rho} \epsilon_{\mu\nu\rho} \left [ B_{\mu\nu}({\bf n}+\hat\rho) - B_{\mu\nu}({\bf n}) \right]
=  Q \delta_{n_1,\frac{L}{2}} \delta_{n_2,\frac{L}{2}} \left [ \delta_{n_3,0} - \delta_{n_3,s} \right].
\ee
As is well known, we cannot find a set of gauge fields, $\theta_\mu({\bf
n})$, that realizes the above set of plaquette values as their field
strength.  Instead, one can find a set of gauge fields that
minimizes the non-compact action in the presence of a flux background,
$B_{\mu\nu}$, given by
\be
S_g =  \sum_{\bf n} \sum_{\mu < \nu=1}^3 \left [ F_{\mu\nu}({\bf n}) -  B_{\mu\nu}({\bf n}) \right]^2;\qquad
F_{\mu\nu}({\bf n}) = \theta_\mu({\bf n}) + \theta_\nu({\bf n}+\hat\mu ) - \theta_\mu({\bf n}+\hat\nu) - \theta_\nu({\bf n}).
\label{gaction}
\ee
The minimum is easily found by going to the momentum space ${\bf k}=(k_1,k_2,k_3)$ for integer $k_\mu$, 
and the solution is given by
\be
\theta_\mu({\bf n}) = \sum_{\bf k} \tilde \theta_\mu({\bf k}) e^{i\frac{2\pi {\bf k}\cdot{\bf n}}{L}};\qquad \tilde\theta_\mu({\bf k}) = \frac{\tilde J_\mu({\bf k})}{\hat k^2};\qquad \hat k^2 = 4 \sum_{\mu=1}^2 \sin^2 \frac{\pi k_\mu}{L},\label{thetak}
\ee
where the current is given by
\be
J_\mu({\bf n})=
\sum_{\nu} \left[ B_{\mu\nu}({\bf n}) -B_{\mu\nu}({\bf n}-\hat\nu)\right];\qquad
\tilde J_\mu({\bf k}) = \frac{1}{L^3} \sum_{\bf n}  J_\mu ({\bf n}) e^{-i\frac{2\pi {\bf k}\cdot{\bf n}}{L}}
.\label{lateqn}
\ee
The current has no zero momentum component and the conservation of the 
current on the lattice is given by
$\sum_\mu \left [ J_\mu({\bf n}) - J_\mu({\bf n}-\hat\mu) \right] =0$.

\subsection{Parity invariance of the field}

Using the field ${\cal A}$ from a continuum Dirac-Monopole pair, it is easy to show that the 
field is parity-invariant under ${\bf x}\to -{\bf x}$ about the mid-point of the Dirac string 
connecting the monopole and anti-monopole. In order to demonstrate this for the background field as defined above, 
let us first define the parity operator $P$ via its action ${\bf n}\to {\bf n}^p={\bf L}-{\bf n}$, where ${\bf L}=(L,L,L)$.
The action of parity on gauge fields on the lattice is then
\be
(P\theta)_\mu({\bf n}) = \theta_\mu^p({\bf n})= -\theta_\mu({\bf L}-{\bf n}-\mu).\label{pardef}
\ee
and the plaquette defined in \eqn{gaction} satisfies
\be
F^p_{\mu\nu}({\bf n}) = F_{\mu\nu}( {\bf L} - {\bf n} - \hat\mu-\hat\nu).
\ee
Under this relation,
the background flux defined in \eqn{diracs} satisfies the property
\be
B^p_{\mu\nu}({\bf n}-{\bf t}) = B_{\mu\nu}({\bf n});\qquad {\bf t} = (-1,-1,s+1-L).
\ee
Therefore, the background field that minimizes, \eqn{gaction}
will satisfy the property
\be
\theta^p_\mu({\bf n} - {\bf t}) = \theta_\mu({\bf n}).
\ee
Let us define the special translation operator $\tau_{\bf t}$ by
\be
\left [ \tau_{\bf t} \psi\right] ({\bf n}) = \psi({\bf n}+{\bf t})\label{spectran}
\ee
and the standard covariant translation operator $T_\mu$ by
\be
(T^\theta_\mu\psi)({\bf n}) = e^{i\theta_\mu({\bf n})} \psi({\bf n}+\hat\mu).\label{tmudef}
\ee
Since
\be
PT^\theta_\mu P = {T^{\theta^p}}^\dagger_\mu\quad {\rm and}\quad \tau^\dagger_{\bf t} T_\mu^{\theta^p} \tau_{\bf t} = T_\mu^\theta,
\ee
we arrive at
\be
{T_\mu^\theta }^\dagger = \bar P T_\mu^\theta \bar P^\dagger;\qquad \bar P = P\tau_{\bf t};\qquad \bar P^\dagger \bar P = \mathbf{I}.\label{modpar}
\ee

\subsection{Defining continuum limit of the background field}

The continuum limit of a lattice field theory is a subtle limit along the lines of constant physics 
near a fixed point of the lattice theory. However, in this paper we consider a comparatively 
trivial continuum limit --- it is possible to define a continuum limit of a background gauge 
field in such a way that length scales associated with the background field remain fixed with 
respect to the lattice size. In other words, we set the physical size of the periodic box to be 
unity by definition and measure all other length scales with respect to it, in which case the 
lattice spacing is $1/L$. For example, we can consider a wave-like lattice gauge field 
$\theta^{\rm wave}_\mu(x)=c_\mu/L\cos(2\pi/L)$ whose continuum limit $L\to\infty$ 
is taken at fixed value of parameter $c_\mu$.
In the case of the monopole-anti-monopole pair, the associated length scale is the lattice distance $s$ between the 
monopole and anti-monopole. Therefore, we define the continuum limit as the $L\to\infty$ limit at a fixed value of 
$f=s/L$. In this paper, we set $f=1/4$. Now, it makes sense to ask whether different lattice discretization 
of the continuum Dirac operator give universal results in the above defined continuum limit.

It is possible to demonstrate the non-trivial nature of the 
monopole background that is discretized on the lattice by using the spherical Dirac monopole field ${\cal A}$.
Since ${\cal A}$ is scale invariant, it easy to see that the corresponding lattice field $\overline{\theta}_\mu({\bf n})$
that connects the lattice site ${\bf n}$ to ${\bf n}+\hat\mu$ remains invariant at fixed ${\bf n}$ for all values of $L$ under the 
above continuum limit. The reason is the following ---
when the lattice spacing is reduced by a factor $k$, the physical distance of a lattice site from the 
monopole reduces 
by a factor $k$ and hence the physical gauge field at the lattice site 
increases by a factor $1/k$. When integrated over a lattice spacing to obtain $\overline{\theta}$, the 
factor $k$ gets cancelled. This is unlike the smooth background $\theta^{\rm wave}$ considered above which 
approaches zero as $1/L$ in the continuum limit.

\bef
\centering
\includegraphics[scale=0.55]{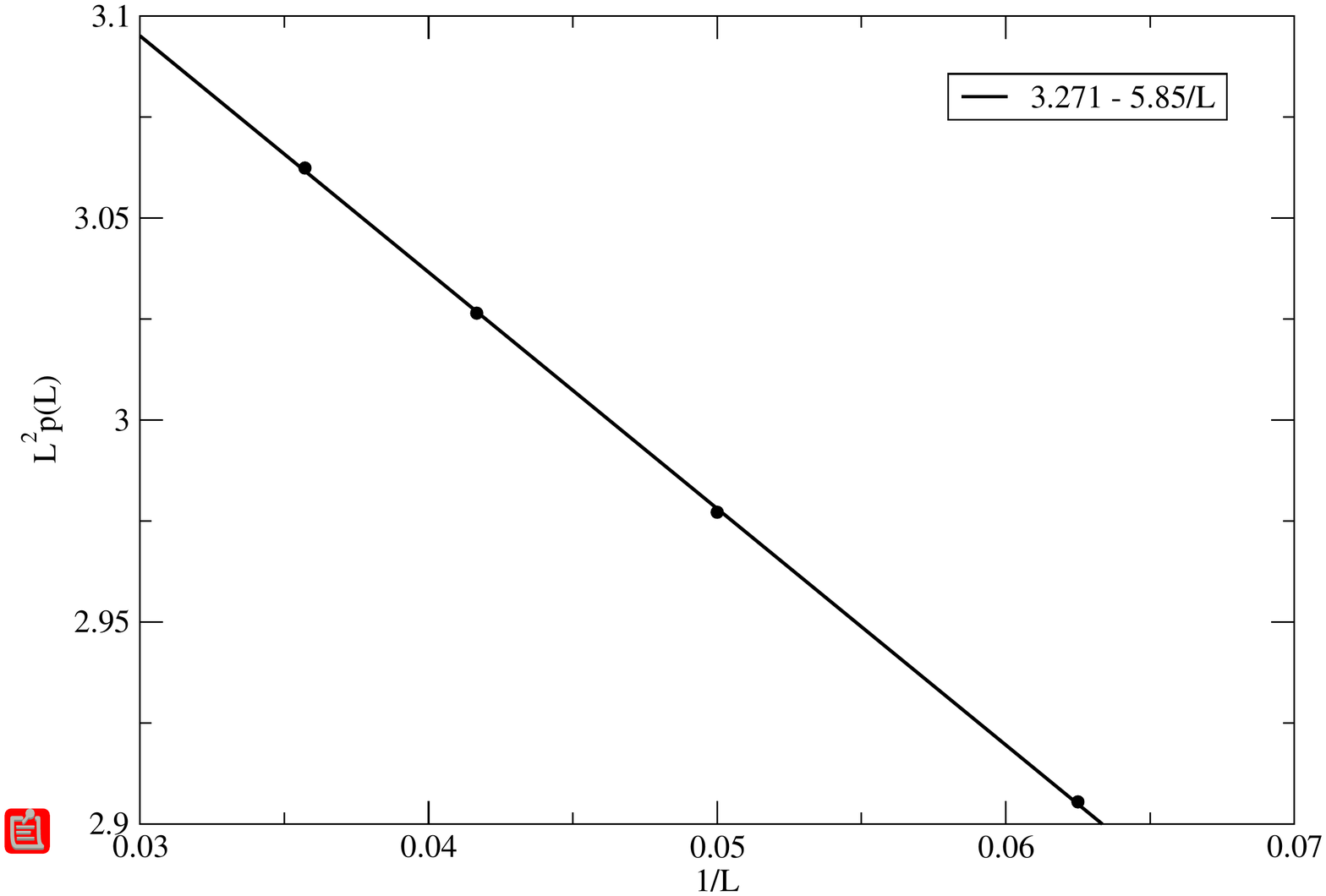}
\caption{The action of the background gauge field as a function of $L$.}
\eef{action}

The lattice-like nature of the background field even in the $L\to\infty$ limit can be seen in 
the scaling of non-compact action $S_b=\sum_{\bf n}\sum_{\mu>\nu} F_{\mu\nu}({\bf n})^2$ with 
$L$ for $F_{\mu\nu}$ obtain through the minimization of \eqn{gaction}. 
The background field does not have a continuum limit in the usual sense where we expect $\theta_\mu({\bf n})$ to be of order $\frac{1}{L}$ and the derivatives
to be order $\frac{1}{L}$. In that case, the average value of the action, namely,
\be
p(L) = \frac{1}{3L^3} S_b(L)
\ee
is expected to go like $\frac{1}{L^4}$. Instead, we find that
\be
p(L) = 3.271\frac{1}{L^2} - 5.85 \frac{1}{L^3}
\ee
for the background field discussed in \scn{maml} with $Q=1$
as shown in \fgn{action}.  This atypical behavior is expected due
to the presence of a monopole-anti-monopole pair in the background
gauge field corresponding to singularities in the flux distribution.
In the following sections, we will study the effect of this on the
low lying spectrum of fermions.

\subsection{Parity-doubling of continuum Dirac spectrum as reference for lattice fermions}

In order to investigate the effect of the singular nature of the monopole-anti-monopole background 
field on lattice regulated fermions, we need to choose an appropriate observable that is characteristic of 
the field and has well a defined property in the unregulated continuum Dirac operator. As we noted above,
a characteristic feature of the background field is its parity invariance. For the continuum 
Dirac operator,
\beq
P \slashed{D}^{\rm cont}(A) P = - \slashed{D}^{\rm cont}(A^p);\qquad \slashed{D}^{\rm cont}(A)=\slashed{\partial}+i\slashed{A},
\eeq{continuumd}
with $A^p_\mu({\bf x})=-A_\mu(-{\bf x})$. For parity invariant fields, $A^p_\mu({\bf x})=A_\mu({\bf x})$ up to
a gauge transformation. This implies the anticommuting relation
\beq
P \slashed{D}^{\rm cont}(A) P = - \slashed{D}^{\rm cont}(A).
\eeq{anticomm}
Since $\slashed{D}^{\rm cont}$ is anti-hermitian, the above anti-commulation property implies that, if
$\psi_+$ is an eigenvector with
\beq
\slashed{D}^{\rm cont}\psi_+ = i \lambda \psi_+,
\eeq{eigcont}
then $\psi_-=P\psi_+$ is an eigenvector with eigenvalue $-i\lambda$. It is convenient to recast this as 
a statement about $\left(\slashed{D}^{\rm cont}\right)^\dagger\slashed{D}^{\rm cont} $:
\beq
\left [ \left(\slashed{D}^{\rm cont}\right)^\dagger \slashed{D}^{\rm cont}  \right] \psi_\pm = \lambda^2 \psi_\pm.
\eeq{ddag}
Thus, there a parity-doubling of eigenvalues of $ \left(\slashed{D}^{\rm cont}\right)^\dagger \slashed{D}^{\rm
cont}$.  As we will
see, the low-lying eigenvalues of $\left(\slashed{D}^{\rm cont}\right)^\dagger \slashed{D}^{\rm cont}$ and their expected
parity-doubling lead to unexpected observations for lattice fermions.

The following will then be our method. We will study the low lying
eigenvalue spectrum of lattice Dirac operators in the limit
$(L,s)\to\infty$ at a fixed $f=\frac{s}{L}$.  Precisely, we will
study the microscopic eigenvalues of the positive definite operator
 $\left(\slashed{D}^{\rm lat}\right)^\dagger \slashed{D}^{\rm lat}$
constructed out of the lattice Dirac operators $\slashed{D}^{\rm
lat}$ for the naive-Dirac, Wilson-Dirac and overlap-Dirac lattice
operators in the above background and analyze the low lying spectrum
as a function of $L$ at  a fixed $Q$ and $f$. We will mainly consider
$Q=1$ and we will set $f=\frac{1}{4}$.  We will work with $L$ that
are multiples of $4$ from $L=12$ to $L=56$. At the end we will study
Wilson-Dirac fermions with $Q=2$ in order to make some conclusions
about the study of compact QED using Wilson-Dirac and overlap-Dirac
fermions.

\section{Naive-Dirac fermions}\label{sec:naive}
\bef
\centering
\includegraphics[scale=0.275]{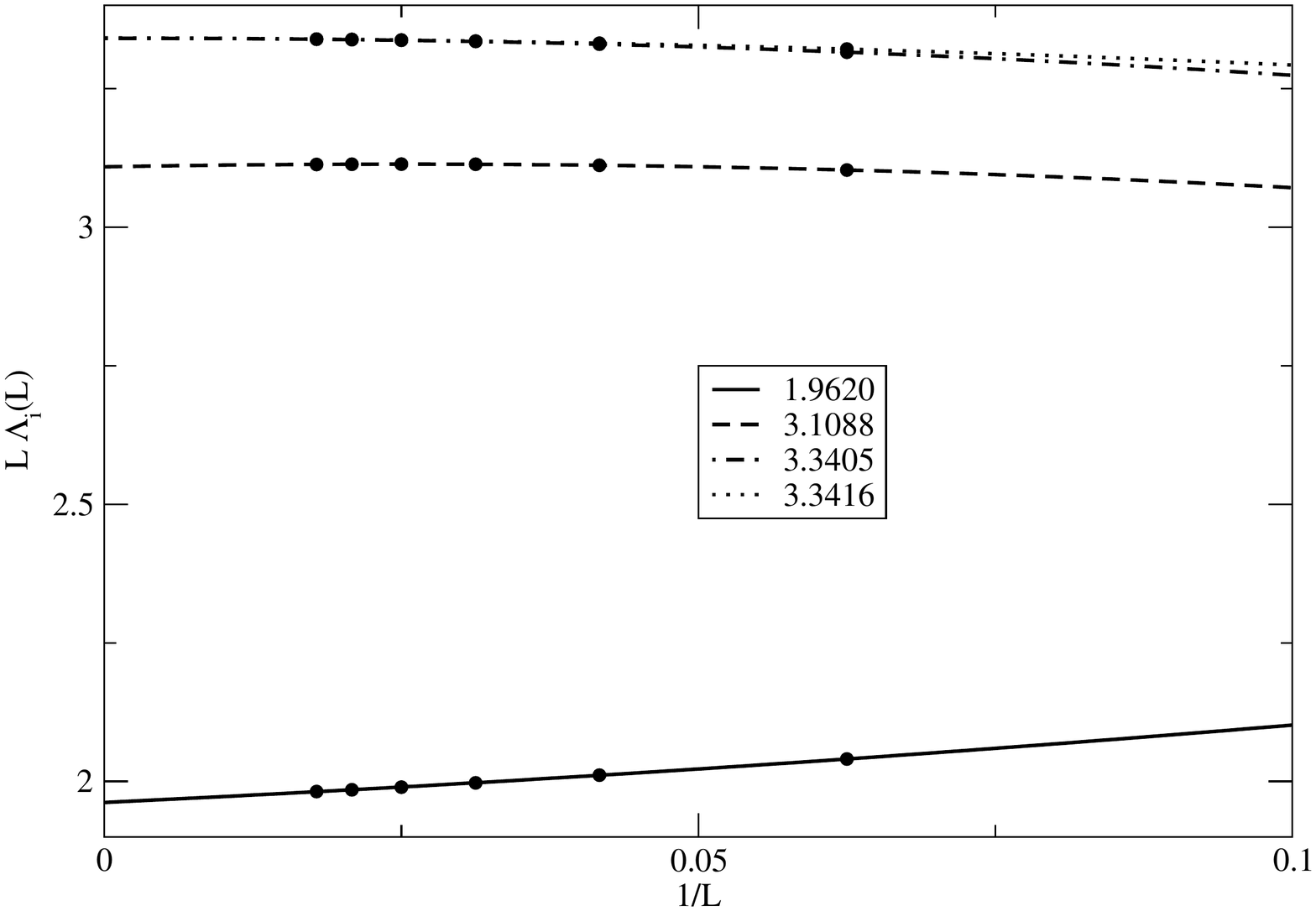}
\includegraphics[scale=0.275]{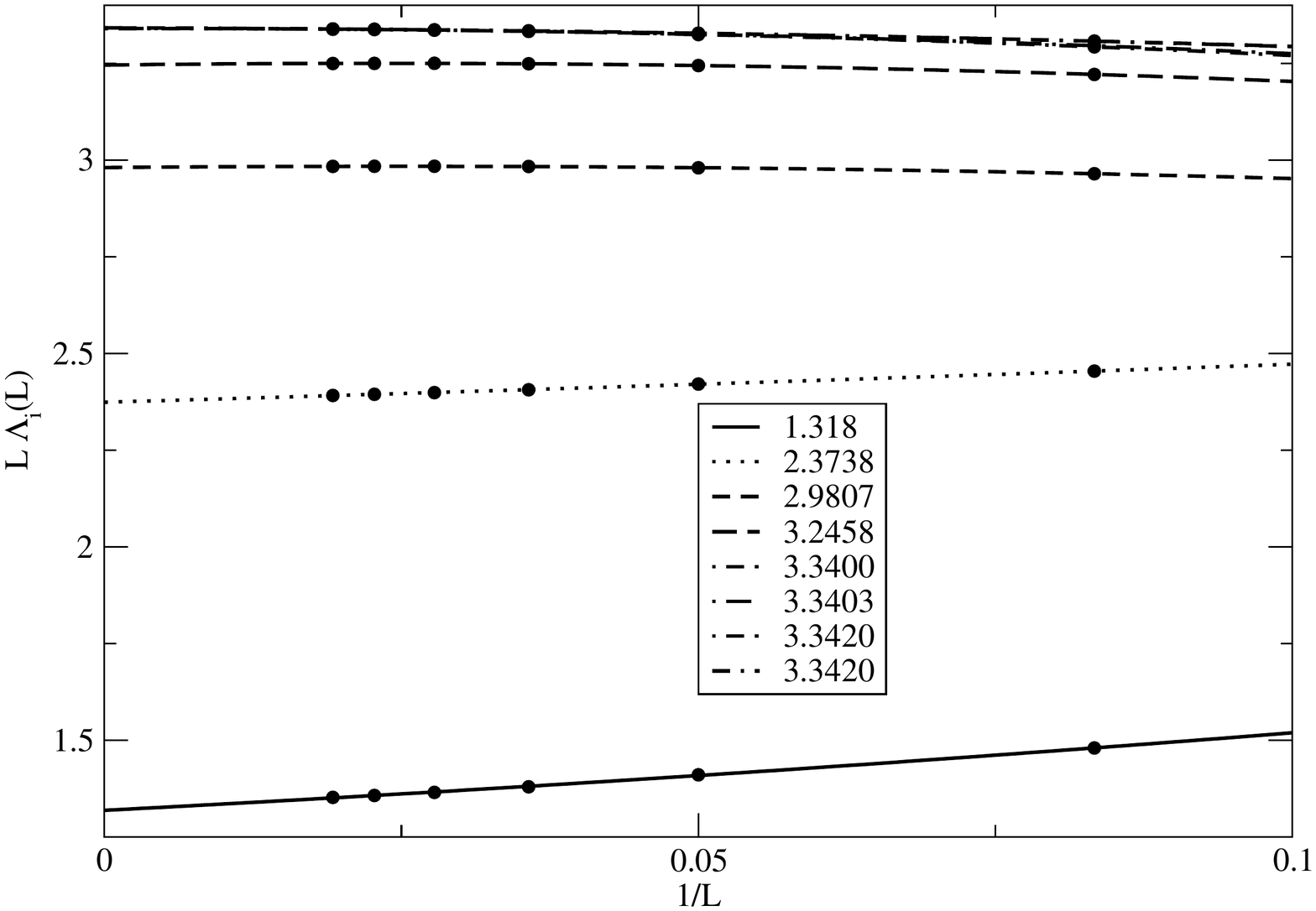}
\includegraphics[scale=0.275]{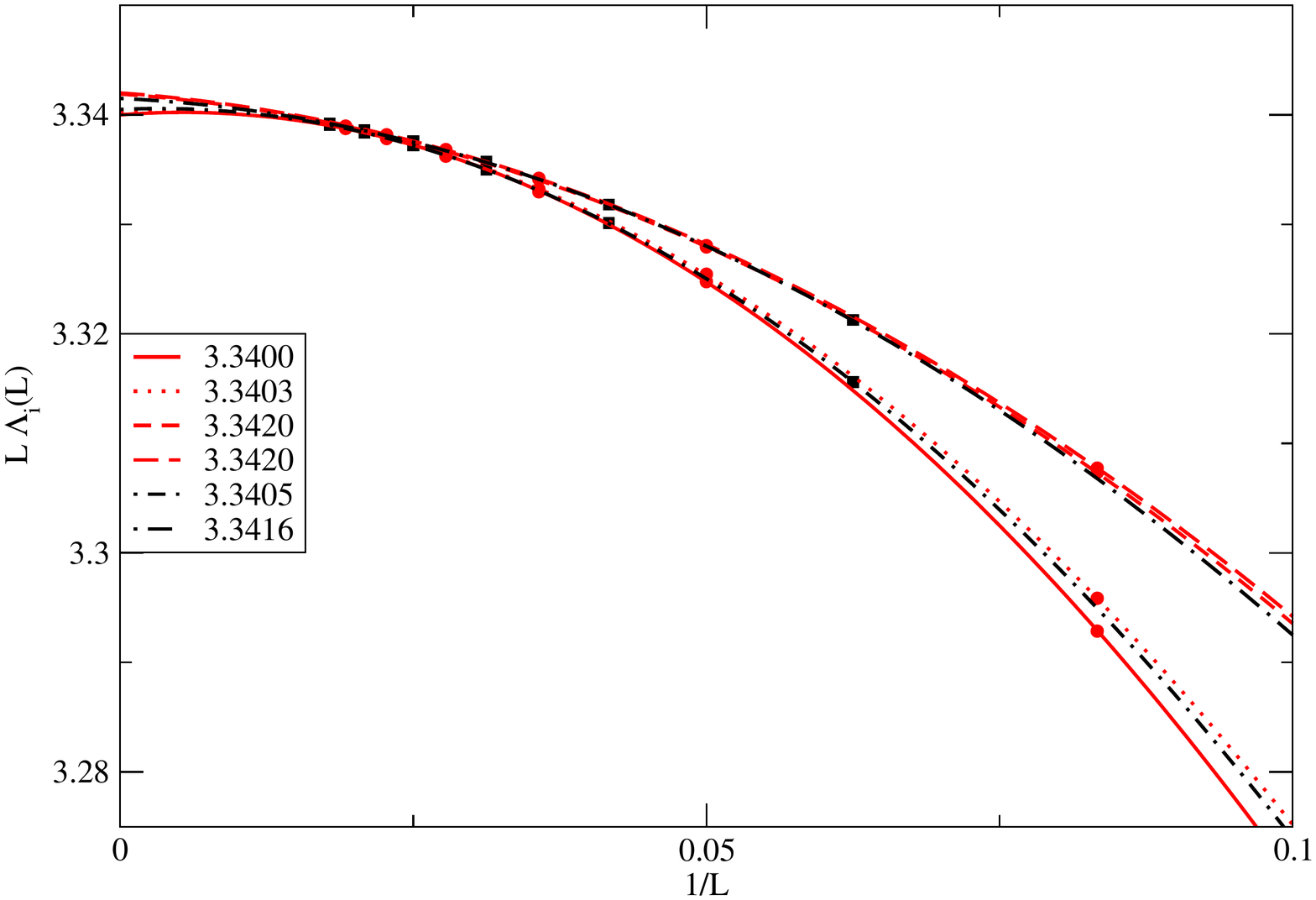}
\caption{The low lying eigenvalues of the naive-Dirac operator as
a function of $L$. The top left plot shows the spectrum for $L=4(2n+2)$; $n=1,2,3,4,5,6$
and show an eight-fold degeneracy.  The top right plot shows the spectrum for $L=4(2n+1)$;
$n=1,2,3,4,5,6$ and show a four-fold degeneracy. The bottom plot shows the third
and fourth distinct eigenvalues for $L=4(2n+2)$ (in black) and the fifth to eighth distinct eigenvalues for
$L=4(2n+1)$ (in red). All these different spectral levels in the bottom panel are expected to become degenerate only when $L\to\infty$.
}
\eef{eigen-naive}
The na\"ive massless Dirac operator in three dimensions is explicitly given by
\be
\slashed{D} = \frac{1}{2} \sum_{\mu=1}^3 \sigma_\mu \left( T_\mu - T_\mu^\dagger\right);\qquad
\left( T_\mu\phi\right)({\bf n}) = e^{i\theta_\mu({\bf n})} \phi({\bf n}+\hat\mu);\qquad T_\mu^\dagger T_\mu = 1;
\qquad
{\slashed{D}}^\dagger = -\slashed{D}.\label{naive}
\ee
This operator is expected to describe a theory with eight degenerate
flavors. Since the staggered-Dirac operator is obtained from the
naive-Dirac operator by a change of basis~\cite{Kogut:1974ag,Gattringer:2010zz},
it is clear that the spectrum will trivially show a two-flavor
degeneracy for all background gauge fields.  In addition, for our
background gauge field that satisfies \eqn{modpar}, we have
a relation similar to the continuum Dirac operator as
\be
\bar P \slashed{D} \bar P^\dagger = -\slashed{D},\label{naivepar}.
\ee
The above parity-doubling will lead to at least a
four-fold degeneracy of  the spectrum of
\be
{\slashed{D}}^\dagger \slashed{D} \psi_i = \Lambda^2_i \psi_i;\qquad 0 < \Lambda_1 < \Lambda_2 < \cdots.
\ee
If naive-Dirac fermions do not break the flavor symmetry, we should
therefore find a sixteen-fold degenerate spectrum.  We will compute
the low lying eigenvalues of ${\slashed{D}}^\dagger \slashed{D}$ using the Ritz
algorithm~\cite{Kalkreuter:1995mm} and impose anti-periodic boundary
conditions in one of three directions (we choose the $y$ direction).
We expect $ \lambda_i =\lim_{L\to\infty} \Lambda_i L $ to be finite
and non-zero. For reference, the three distinct lowest eigenvalues
for free fermions with anti-periodic boundary conditions in one of
three directions will be $(\lambda_1,\lambda_2,\lambda_3)=(1, \sqrt{5},3)\pi$.  The results
for the lowest thirty-two eigenvalues are shown in \fgn{eigen-naive}.
 Let us first focus on the top left plot in \fgn{eigen-naive}
which correspond to even values of $s$ obtained by setting $L=4(2n+2)$
for $n=1,2,3,4,5,6$.  The first two-lying distinct eigenvalues have an
eight-fold degeneracy and the third distinct eigenvalue has an almost
sixteen-fold degeneracy. Therefore, we conclude that the eight-fold
flavor symmetry is broken into two remnant four-fold flavor symmetries
at the lowest level and this effect persists all the way to
$L\to\infty$.  When we look at the spectrum in the top right plot corresponding
to odd values of $s$ obtained by setting $L=4(2n+1)$ for $n=1,2,3,4,5,6$,
we see that the four low lying distinct eigenvalues have only a
four-fold degeneracy. Therefore, the flavor symmetry is broken to the minimum required by the trivial two-fold symmetry
required by the presence of two copies of staggered fermions.
Furthermore, this flavor breaking persists all the way to $L\to\infty$. Focussing on the bottom plot,
 the third and fourth distinct
eigenvalues when $L=4(2n+2)$ and the fifth to eighth
distinct eigenvalues when $L=4(2n+1)$ all approach  a sixteen-fold degeneracy when $L\to\infty$
and  the result
from $L=4(2n+1)$ and $L=4(2n+2)$ match.  We fitted
\be
\Lambda_i L = \lambda_i + \frac{\alpha_i}{L} + \frac{\beta_i}{L^2} \label{evfit}
\ee
using a standard least square fit and the fitted values of $\lambda_i$
 are quoted in \fgn{eigen-naive} as legends of the corresponding
fits. To make the point the sixteen-fold degeneracy is achieved
only when $L\to\infty$, we have listed the fits from the four
four-fold degenerate spectrum for $L=4(2n+1)$ 
and the two eight-fold degenerate spectrum for $L=4(2n+2)$ in all three plots.
The convergence in the actual data as $L\to\infty$ is better than what is seen in the fitted values at $L\to\infty$.
We expect any slight disagreement between the almost degenerate extrapolated eigenvalues to be within
 systematical errors associated with the fit form in \eqn{evfit}.

\section{Wilson-Dirac fermions}\label{sec:wilson}

The Wilson term,
\be
B -m_w =\frac{1}{2} \sum_{\mu=1}^3 \left( 2 - T_\mu - T_\mu^\dagger\right)-m_w;\ \ \ \
B=B^\dagger,\label{wilson}
\ee
will lift the doublers observed in \scn{naive} and
\be
X=B-m_w+\slashed{D};\qquad X^\dagger = B -m_w - \slashed{D}\label{xop}
\ee
are Wilson-Dirac fermions for a pair of two-component fermions
related by parity.  The mass term is parity even as long as we view
$(B-m_w)$ as a whole as the mass term with $m_w\in (-2,2)$.  We
have used an unconventional notation for the sign of the mass to
make it convenient for the definition of overlap-Dirac fermions.

The Wilson-Dirac fermion
action for a pair of two-component fermions that is parity invariant
is given by
\be
S_{\rm fw} = \begin{pmatrix} \bar\phi_2 & \bar\phi_1 \end{pmatrix}
 \begin{pmatrix} 0 & X^\dagger \cr X & 0 \end{pmatrix} \begin{pmatrix} \phi_1 \cr \phi_2 \end{pmatrix}.\label{faction}
\ee 
Fo our particular background which obeys \eqn{modpar}, we have $\bar P^\dagger X \bar P = X^\dagger$, and we can identify
$\phi_2$ with $\bar P^\dagger \phi_1$. Since we can only discuss the spectrum of a four-component parity invariant fermion, we do not have
the double degeneracy present in two-component naive fermions at the expense of removing the doublers.
The eigenvalues of the four-component fermion operator come in $\pm\Lambda_i$ pairs where $\Lambda_i > 0$ are obtained from
the eigenvalue problem
\be
X^\dagger X \psi_i = \Lambda_i^2 \psi_i;\qquad 0 < \Lambda_1 < \Lambda_2 < \cdots \label{ebasis}.
\ee

\bef
\centering
\includegraphics[scale=0.55]{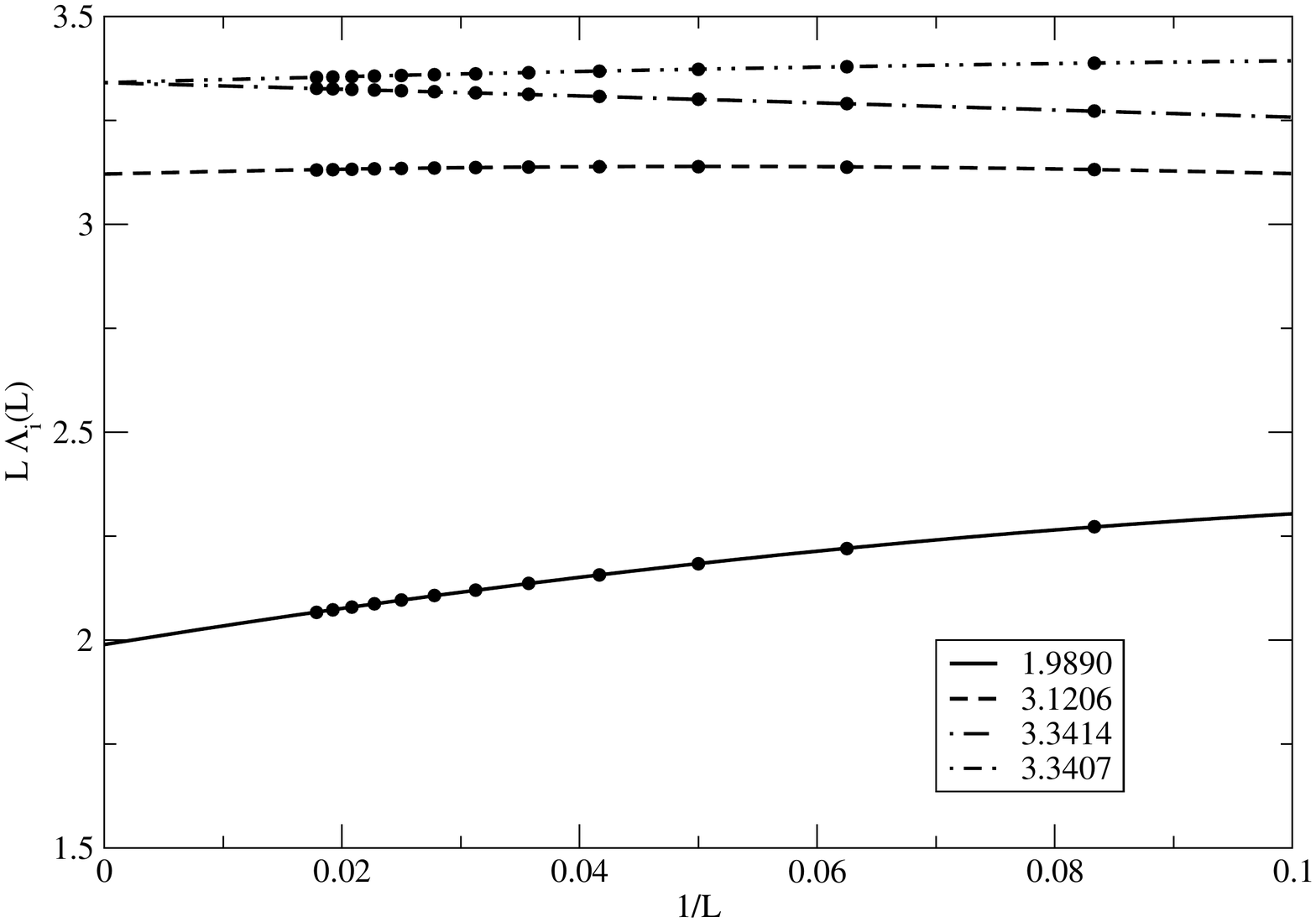}
\caption{The low lying eigenvalues of the Wilson-Dirac operator with $m_w=0$ as a function of $L$. 
}
\eef{eigen-wilson-massless}

Using \eqn{xop}, we can write
\be
X^\dagger X = -{\slashed{D}}^2 + [B,\slashed{D}] +(B-m_w)^2.
\ee
If we consider gauge field configurations generated by the standard
non-compact Wilson action (gauge fields on links will scale as
$\frac{1}{L}$  at a fixed $L$ when the background field is set to
zero in \eqn{gaction}) as was done in~\cite{Karthik:2015sgq}, we
expect $\slashed{D}$ to scale like $\frac{1}{L}$ and $B$ to scale like
$\frac{1}{L^2}$.  To maintain a finite physical mass, we would set
$m_w=\frac{m}{L}$ where we keep $m$ fixed as we take $L\to\infty$.
In this set-up, we expect
\be
\lambda_i (m) = \lim_{L\to \infty}  L\sqrt{\Lambda^2_i  -m_w^2}\label{wilsonmscaled}
\ee
to be finite and non-zero. Furthermore, we expect $\lambda_i(m)$
to be independent of $m$ and consistent with the value obtained
using naive-Dirac fermions.

\subsection{Properties at finite physical mass $m=m_w L$}

We first set $m_w=0$ and plot the four lowest eigenvalues,
$L\Lambda_i(L)$, as a function of $\frac{1}{L}$ in
\fgn{eigen-wilson-massless}. The data fit \eqn{evfit} well and the
fitted values of $\lambda_i$ are quoted in
\fgn{eigen-wilson-massless} as legends of the corresponding fits.
On the one hand, the two lowest eigenvalues approach different
limits as $L\to\infty$ showing that Wilson-Dirac fermions do not
recover a double degenerate spectrum realized by naive fermions
that satisfies \eqn{naivepar}. On the other hand, we see that there
is good agreement in the $L\to\infty$ limit between the two lowest
eigenvalues ($\lambda_1$ and $\lambda_2$) for the Wilson-Dirac
operator and
 the two lowest eigenvalues associated with the black lines
(case of eight-fold degeneracy) in \fgn{eigen-naive}.  The doubling
seen in the sixteen-fold degenerate spectrum of naive-Dirac fermions
in \fgn{eigen-naive} is also seen in \fgn{eigen-wilson-massless},
since $\lambda_3$ and $\lambda_4$ are equal.  Furthermore, the
values for $\lambda_3=\lambda_4$ matches well with the corresponding
value obtained from naive-Dirac fermions.  We conclude that naive-Dirac
and massless Wilson-Dirac fermions behave in the same manner in the continuum limit with
$Q=1$ -- (i) the two lowest eigenvalues show a splitting either due
to breaking of flavor symmetry or due to the need for two different
two-component Wilson-Dirac operators to realize a single fermion
flavor; (ii) the rest of spectrum show the expected two-fold
degeneracy per two-component flavor (explicitly seen  for the third
distinct eigenvalue).

\bef
\centering
\includegraphics[scale=0.55]{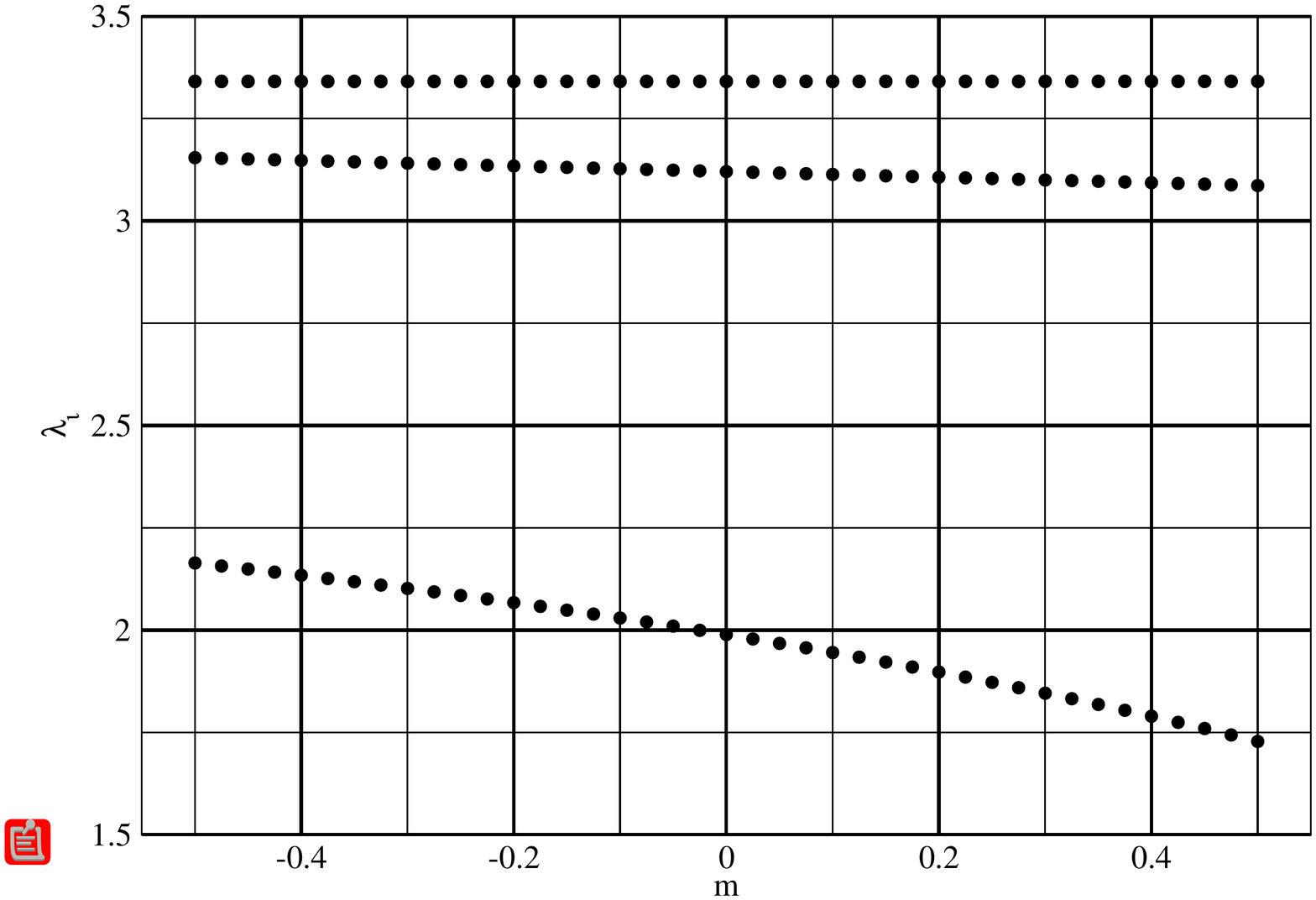}
\caption{The low lying eigenvalues, $\lambda_i(m)$, obtained in the limit of
$L\to\infty$ as a function of $m$.
}
\eef{eigen-wilson}

In order to observe possible effects due the the Wilson term not
being irrelevant, we proceed to study the behavior of the eigenvalues
as a function of $m=m_w L$. To this end, we plot the first four
values of $\lambda_i(m)$, obtained by fitting the right-hand side
of \eqn{wilsonmscaled} using \eqn{evfit}, in \fgn{eigen-wilson}.
We note that $\lambda_1(m)$ and $\lambda_2(m)$ depends on $m$
suggesting that $B$ and $[B,\slashed{D}]$ do not scale naively as expected.
This is an effect of the background as viewed by Wilson-Dirac
fermions. But we see that $\lambda_3(m)=\lambda_4(m)$ are independent
of $m$. The effect of a non-smooth background with $Q=1$ affects
only the two lowest eigenvalues even as a function of $m$.  Note
that naive-Dirac fermions will show the expected quadratic dependence
of mass simply because the mass term commutes with $\slashed{D}$.

\bef
\centering
\includegraphics[scale=0.275]{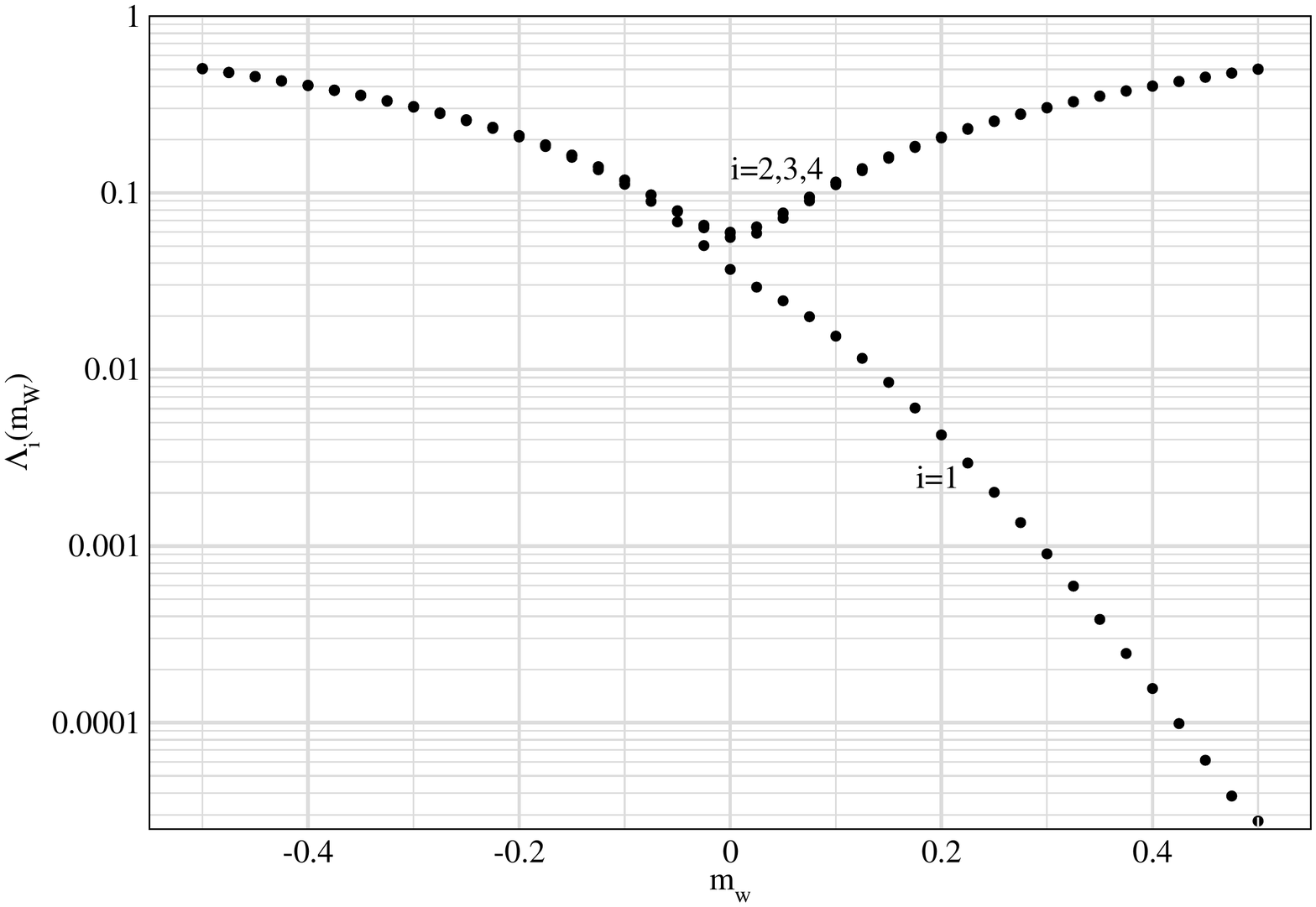}
\includegraphics[scale=0.275]{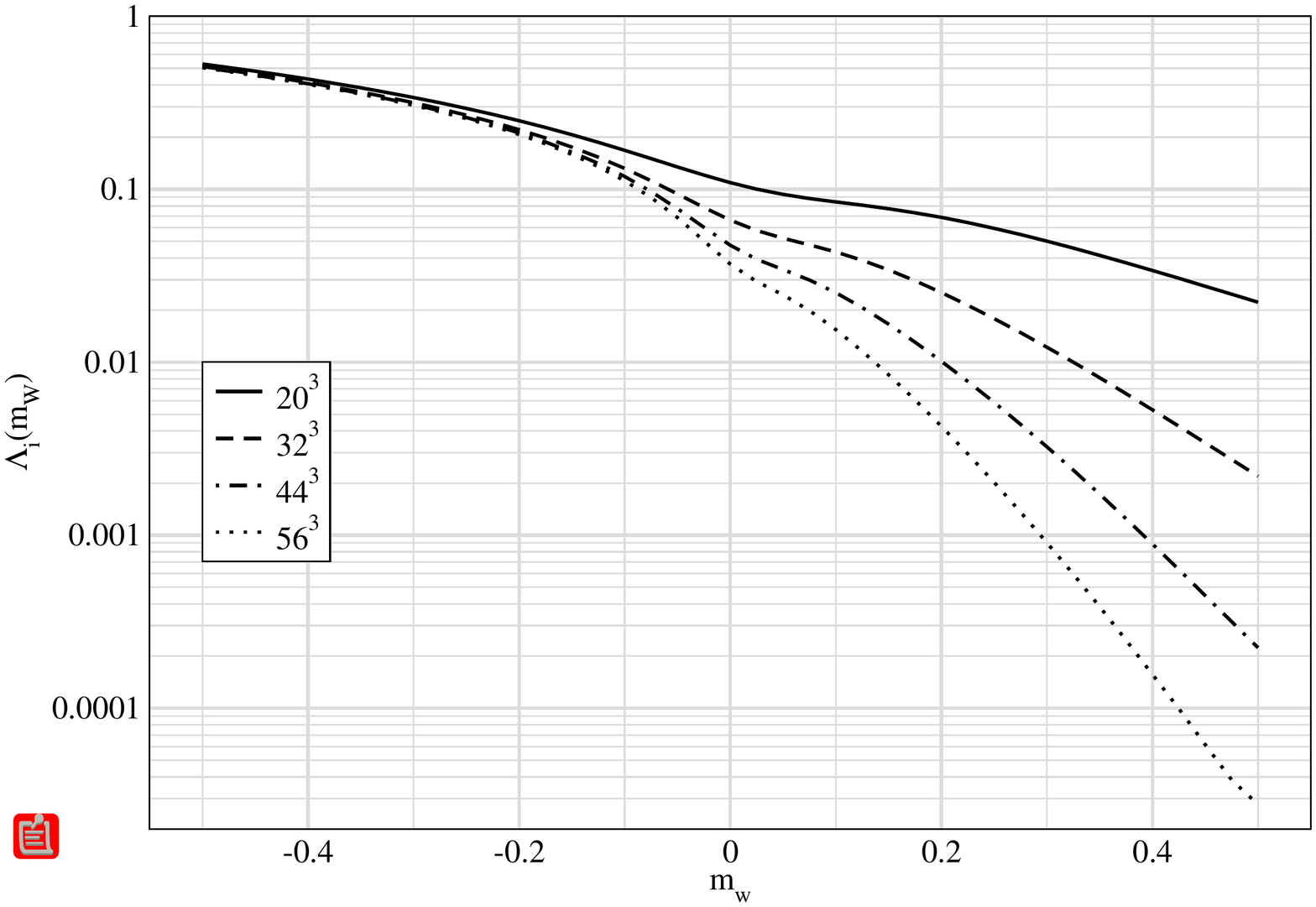}
\caption{The low lying eigenvalues, $\Lambda_i(m_w)$ as a function of $m_w$ are shown for $L=56$ are shown in the
left panel.
The lowest eigenvalue that behaves anomalously for $m_w > 0$ is shown for four different values of $L$ in the
right panel.
}
\eef{wilson-flow-q1}

\subsection{Properties at Wilson mass $m_w$ that is relevant to the kernel of overlap operator}

Finally, we need to understand the behavior of the low lying
eigenvalues as a function of $m_w$ when it is kept fixed as we vary
$L$. As long as $m_w\ne 0$, it corresponds to a fermion with infinite
mass that appears as a kernel for the overlap-Dirac operator.  A
plot of the four low lying eigenvalues, $\Lambda_i(m_w)$ is shown
in the left panel of \fgn{wilson-flow-q1} for $L=56$ and the effect of a background that is not
continuum-like is evident in the behavior of the lowest eigenvalue.
The higher eigenvalues seem to show a behavior that reaches a minimum
at $m_w=0$. The lowest eigenvalue on the other hand shows two distinct behaviors for 
$m_w<0$ and $m_w>0$. The right panel of \fgn{wilson-flow-q1} shows that the lowest eigenvalue at a fixed $m_w$ decreases with increasing $L$ for $m_w > 0$ whereas the lowest eigenvalue approaches a non-zero limit at infinite $L$ for $m_w < 0$. For $m_w<0$, the eigenvalue $\Lambda_1$ at a fixed $m_w$ 
approaches $m_w$ in the $L\to\infty$ limit, with finite $L$ corrections that are polynomial in $1/L$.
This is similar to the behavior seen in the higher eigenvalues as well. This is shown for a 
fixed value $m_w=-0.275$ in the top-left panel of \fgn{eigen-wilson-mw} where $\Lambda_1^2$ is 
plotted as a function of $1/L$. For $m_w>0$, the lowest eigenvalue approaches zero with a distinct 
$\exp\left(-\beta(m_w) L\right)$ behavior for larger $L$ with a $m_w$ dependent coefficient $\beta(m_w)$.
This is demonstrated for $m_w=0.275$ in \fgn{wilson-flow-q1} by plotting $\log(\lambda_1^2)$ as a function 
of $L$ where we observe a good description of the large $L$ data by a simple $\exp\left(-\beta(0.275)L\right)$ shown by 
the line. On the other hand, the higher eigenvalues are gapped at finite $m_w >0$ for $L\to\infty$ 
as we would naively expect. If we examine the dependence of the $\beta(m_w)$ as a function of $m_w$,
we find $\beta(0)$ is consistent with zero and increases with $m_w$ as shown in the bottom 
panel of \fgn{wilson-flow-q1}.

We need to study the consequence of the above anomalous behavior of the lowest eigenvalue
on the overlap-Dirac operator spectrum where $m_w > 0$ only plays
the role of a regulator and one expects physics to be independent
of the choice of $m_w$. In addition, the presence of
 the one anomalously low lying eigenvalue for positive $m_w$
will affect the numerical computation using the overlap-Dirac
operator.
\bef
\centering
\includegraphics[scale=0.275]{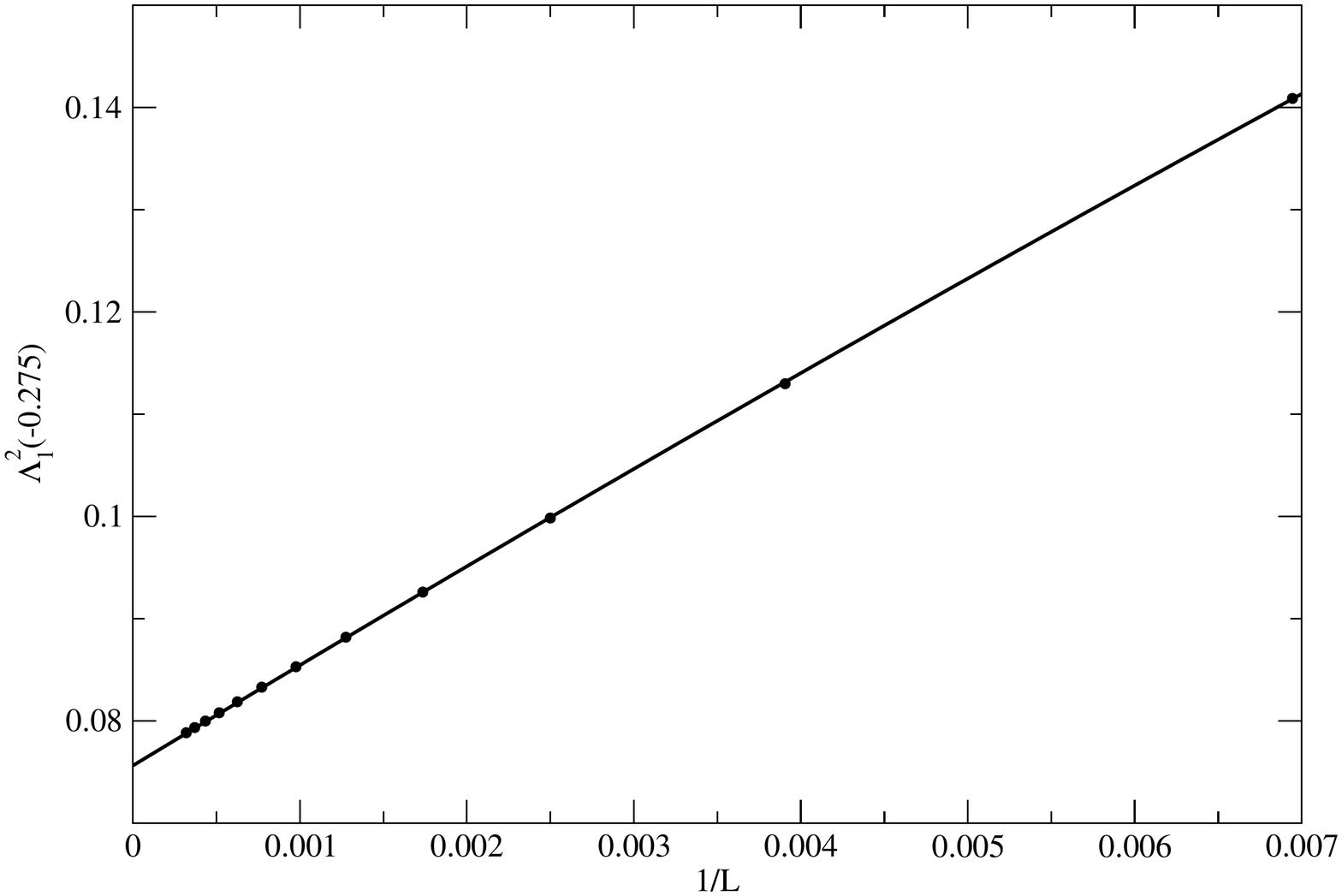}
\includegraphics[scale=0.275]{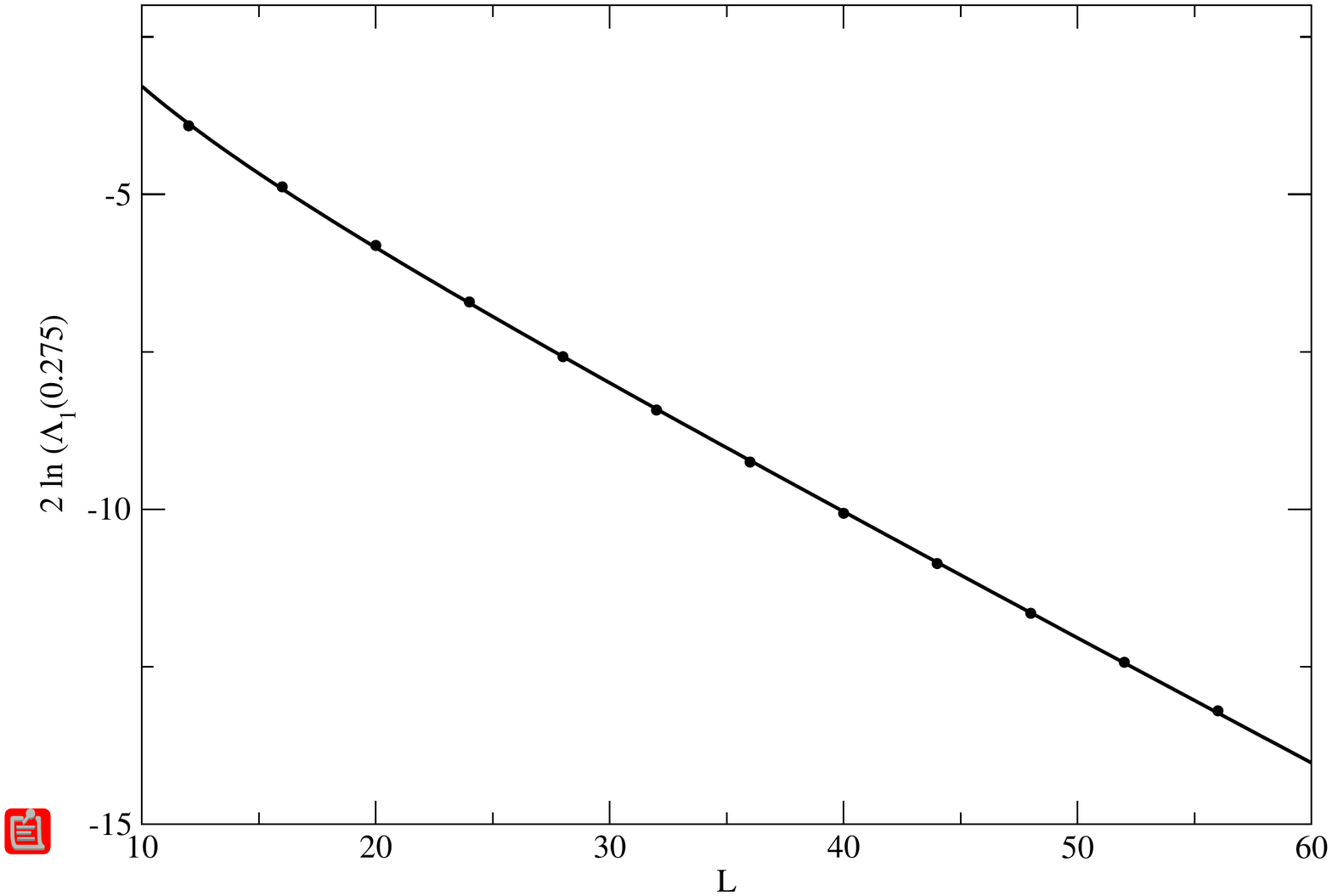}
\includegraphics[scale=0.275]{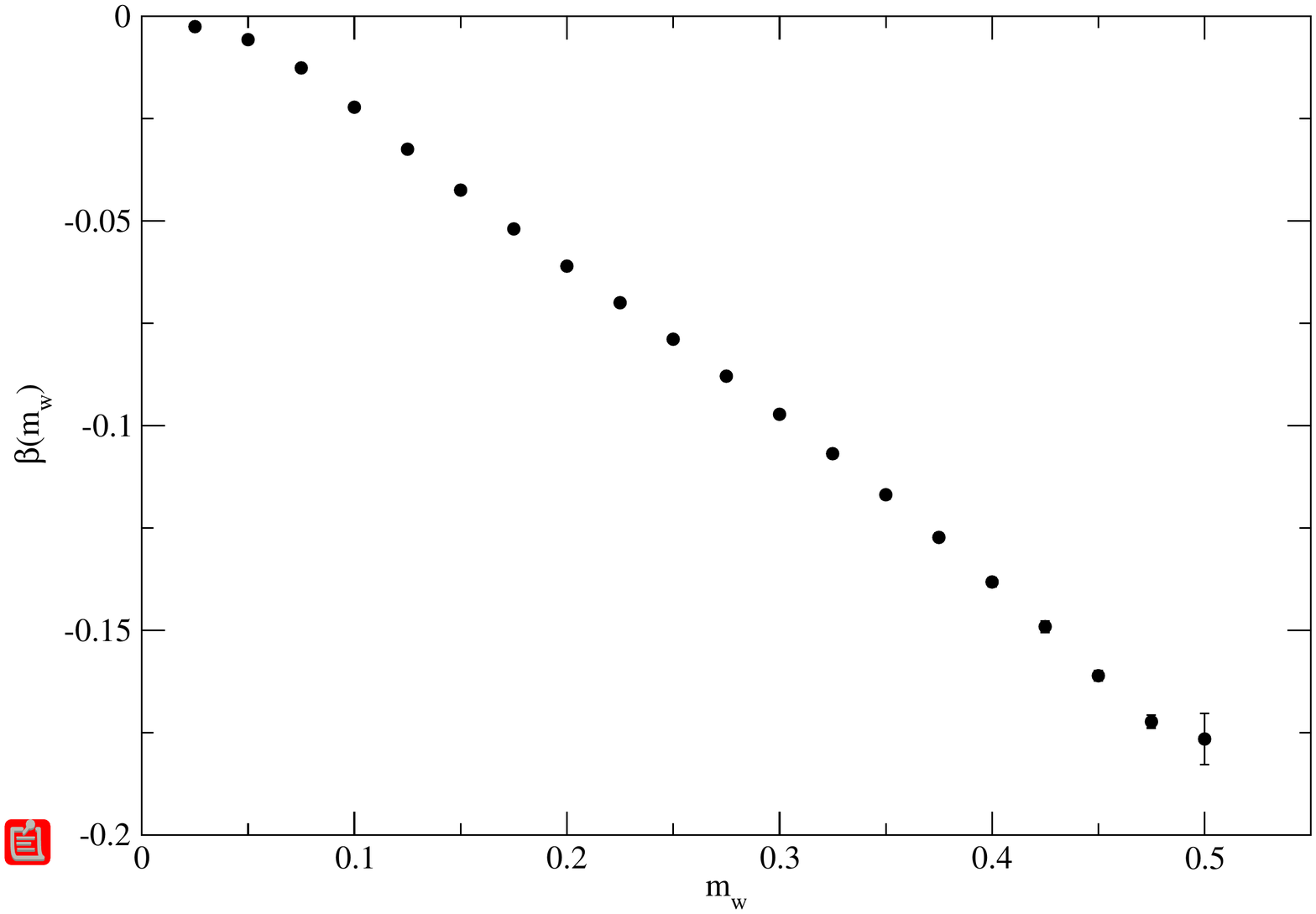}
\caption{
In the top-left panel, the approach of $\Lambda^2_1$ to $m_w^2$ is shown as a function 
of $1/L$ for $m_w=-0.275$. In the top-right panel, the exponential decrease of $\Lambda^2_1$ with 
increase in $L$ is shown for $m_w=0.275$. In the bottom panel, the $m_w$ dependence of $\beta(m_w)$ for the asymptotic exponential
decrease $\exp(-\beta(m_w) L)$ for $m_w>0$ is shown.}
\eef{eigen-wilson-mw}

\section{Overlap-Dirac fermions}

\bef
\centering
\includegraphics[scale=0.55]{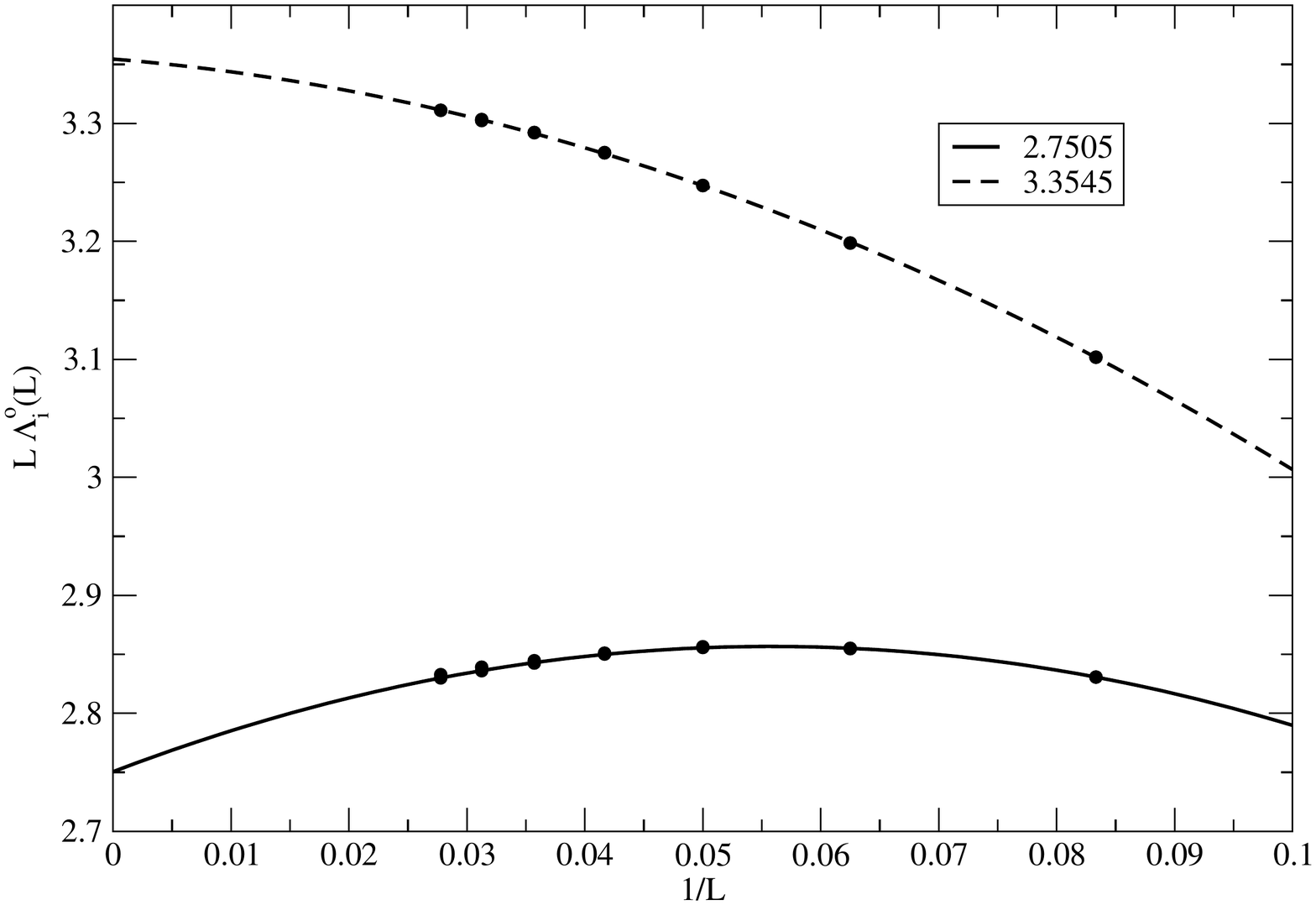}
\caption{The two low lying distinct eigenvalues, $\Lambda_i^o$ as a function of $L$.
}
\eef{eigen-overlap}

The two different two component massless overlap-Dirac operators are 
\be
D_o = \frac{1+V}{2};\qquad {\rm or}\qquad D^\dagger_o = \frac{1+V^\dagger}{2};\qquad V = X\frac{1}{\sqrt{X^\dagger X}}.
\ee
Whereas the presence of the Wilson term in the Wilson-Dirac operator spoiled the commutativity of $X$ and $X^\dagger$,
 $D_o$ commutes
with $D_o^\dagger$. In that sense, overlap-Dirac operator is closer to a continuum Dirac operator -- $D_o$ cannot be anti-hermitian since it has to correctly reproduce the parity anomaly. Since our background field satisfies \eqn{modpar}
 the spectrum of $V$ has the following property that results in a double degeneracy in the spectrum of $D_o^\dagger D_o$.
Since
\be
\left [ \bar P^\dagger V \bar P \right] = V^\dagger,\label{parV}
\ee
we have
\be
V \psi_j = e^{i\phi_j} \psi_j \quad \Rightarrow   V \left [ \bar P\psi_j\right]
= e^{-i\phi_j} \left[ \bar P \psi_j\right],\label{overpar}
\ee
which will result in a double degeneracy in the spectrum of
\be
D_o^\dagger D_o = \frac{2+V+V^\dagger}{4}.
\ee

The analysis in \scn{wilson}
has shown the presence of an anomalously small eigenvalue of $X^\dagger X$ for $m_w > 0$. 
The mass, $m_w$, acts as a regulator for overlap-Dirac fermions and therefore
it is natural to study the spectrum of $D_o^\dagger D_o$ as a function of $m_w$.
 Algorithmically, one uses a rational approximation~\cite{vandenEshof:2002ms,Chiu:2002eh} of the type
\be
\frac{1}{\sqrt{X^\dagger X}} = \sum_{i=1}^n \frac{r_i}{X^\dagger X + p_i}
\ee
where the values of the residues, poles and the number of them are chosen to approximate the operator on the left-hand side to a desired accuracy
in the needed range. This range always has a lower limit away from zero and the presence of a very small eigenvalue of $X^\dagger X$ has to be taken
care of by performing
\be
\frac{1}{\sqrt{X^\dagger X} }{\bf v} = \frac{1}{\sqrt{\Lambda_1}} \left({\bf w}_1^\dagger {\bf v} \right) {\bf w}_1 +
\sum_{i=1}^n \frac{r_i}{X^\dagger X + p_i} \left(1- {\bf w}_1{\bf w}_1^\dagger\right) {\bf v};\qquad X^\dagger X w_1 = \Lambda_1 w_1.
\ee
With this algorithm in place for numerically dealing with the overlap-Dirac operator, we computed the four low lying eigenvalues of 
\be
\left[ D_o^\dagger D_o \right] \psi_i = \left[ \frac{\Lambda_i^o}{2m_w} \right]^2 \psi_i,
\ee
where we have accounted for the trivial mass renormalization that arises from the mass of the Wilson-Dirac fermion~\cite{Edwards:1998wx}. Due to the fact that the lowest eigenvalue of the Wilson-Dirac operator becomes
very small as $L$ is increased, we only went up to $L=36$ where the lowest eigenvalue is still large enough to
enable its projection to the desired accuracy.
The spectrum clearly comes in degenerate pairs due to \eqn{overpar}. The approach to the infinite $L$ limit of the two low-lying distinct eigenvalues,
$\Lambda_i^o$,
is shown in \fgn{eigen-overlap} with $m_w=0.425$ where we fitted the data to the form like for naive-Dirac fermions, namely,
as in \eqn{evfit}.
If we compare with the result for Wilson-Dirac fermions in \fgn{eigen-wilson-massless}, we see that there is a reasonable agreement between the second distinct eigenvalue of the massless overlap-Dirac
operator and the third distinct eigenvalue of the massless Wilson-Dirac operator that is doubly degenerate.
The lowest eigenvalue of the overlap-Dirac operator that also shows a double degeneracy falls in between the two lowest eigenvalues of the Wilson-Dirac operator and it shows strong finite $L$ effects but there is no simple relationship between the lowest eigenvalue of the overlap-Dirac operator and the two lowest eigenvalues
of the Wilson-Dirac operator.

Finally we plot the spectrum of the two low lying distinct eigenvalues of the massless overlap-Dirac operator as a function
of the Wilson-Dirac mass in \fgn{eigen-overlap-mw}. Two features are evident. There is clear evidence of a double degeneracy
in the spectrum within numerical errors arising from the anomalously small eigenvalue of $X^\dagger X$ being not treated accurately enough.  The spectrum is essentially independent of $m_w$ for $m_w > 0.3$.
If the background configuration was continuum like, we would have seen an independence on $m_w$
over the entire range.

\bef
\centering
\includegraphics[scale=0.55]{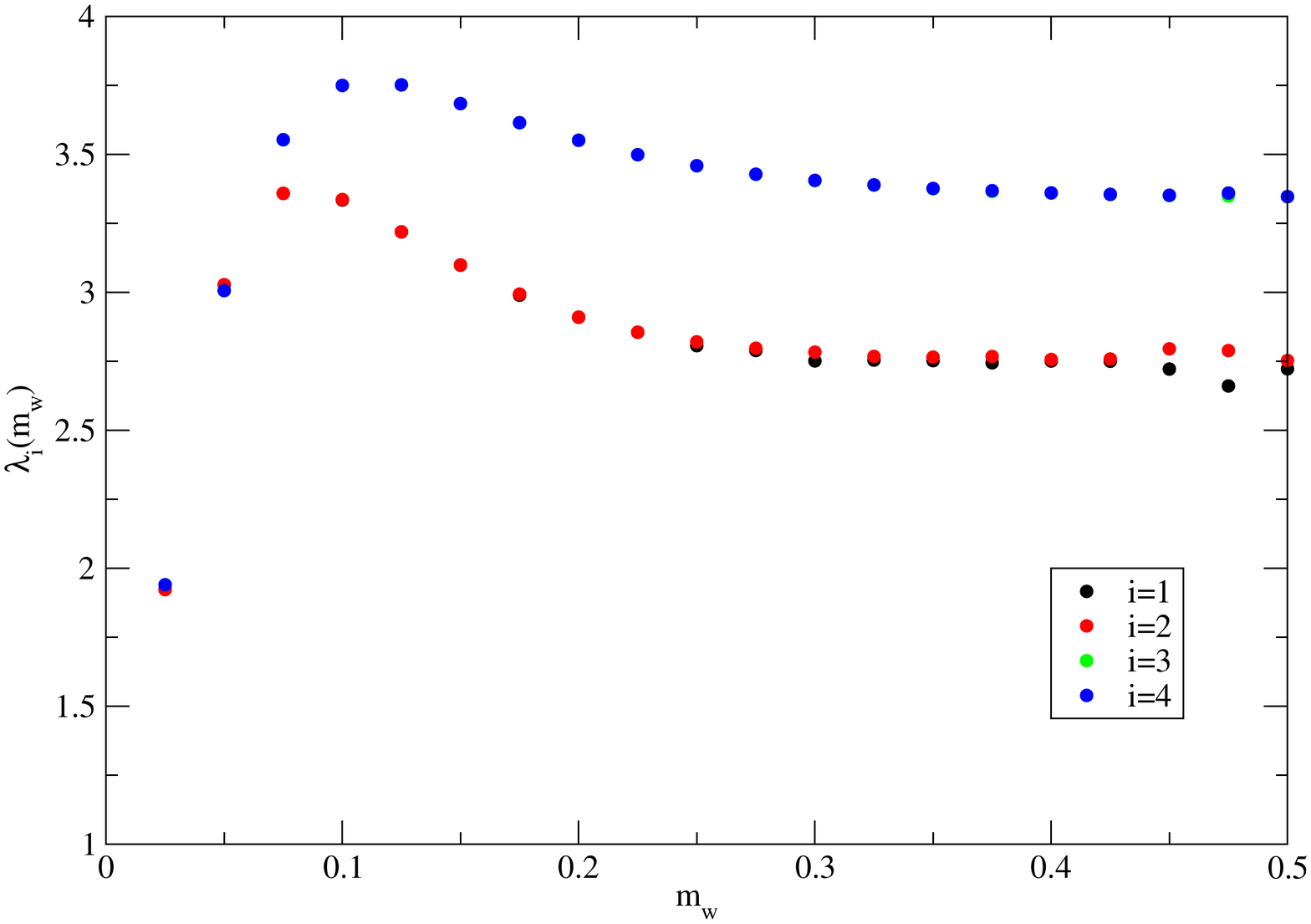}
\caption{The two low lying distinct eigenvalues, $\lambda_i^o$ as a function of $m_w$.
}
\eef{eigen-overlap-mw}

\section{Conclusions}

We defined a background flux corresponding to a monopole-anti-monopole pair separated by a distance $\frac{L}{4}$ on a $L^3$ lattice
by a non-compact flux of $2\pi$ units on a single plaquette in the $z$ direction for an extent of $\frac{L}{4}$.
Using the standard non-compact Wilson action on the lattice, we found the non-compact link variables that minimizes the action in the
presence of the above background. 
A standard continuum limit does not exist for the gauge field that minimizes the action -- the non-compact
link variables do not approach zero as we take $L\to\infty$. This is akin to discretizing a spherical monopole -- the link variables on the plaquette
surrounding the monopole do not go to zero as we take $L\to\infty$. The main question we asked in this paper is the following: Let us couple
the monopole-anti-monopole background to a parity invariant lattice massless fermion action using the compact link variables. 
How do different versions of lattice regularization show the effect of a background that is not continuum like?

Due to the background gauge field being invariant under a combination of parity and a particular lattice translation given by \eqn{modpar}
we expect the spectrum to be doubly degenerate if the lattice fermion is able to respect this symmetry. Naive-Dirac fermion respects this symmetry
but describes eight (four if we reduced it to staggered-fermions) fermion flavors. Wilson-Dirac fermion does not respect this symmetry because the
doublers are lifted by realizing
 the two different
two-component fermions related by parity  by an operator and its hermitean-conjugate that do not commute.
As such neither naive-Dirac fermion nor Wilson-Dirac fermion show a doubly degenerate spectrum at the lowest level for $Q=1$: the sixteen-fold degeneracy
for eight flavors of naive-Dirac fermions is either split into two eight-fold or four four-fold degeneracies implying that flavor symmetry is not realized even when $L\to\infty$; the two-fold degeneracy for one flavor of Wilson-Dirac fermion is split into two implying that Wilson-Dirac fermion does not recover the expected degeneracy even when $L\to\infty$. In spite of this, the spectrum of naive-Dirac fermions and massless Wilson-Dirac fermions match well.
The effect of splitting of the lowest two-fold degenerate level is also seen in the two lowest eigenvalues of the spectrum of the Wilson-Dirac operator with a physically finite mass.
In addition to this unanticipated behavior, Wilson-Dirac fermion has an anomalously small eigenvalue for one sign of the Wilson-Dirac mass that realizes a non-zero Chern-Simons term~\cite{Coste:1989wf,Karthik:2015sza}.  Contrary to Wilson-Dirac fermions, the low lying eigenvalues 
of the overlap-Dirac show the anticipated two-fold degeneracy as long
as we have evaluated the action of the overlap-Dirac operator accurately. The spectrum is independent of the Wilson-Dirac mass parameter that appears in the kernel of the overlap-Dirac operator as long as the Wilson-Dirac mass parameter is away from zero.

\bef
\centering
\includegraphics[scale=0.275]{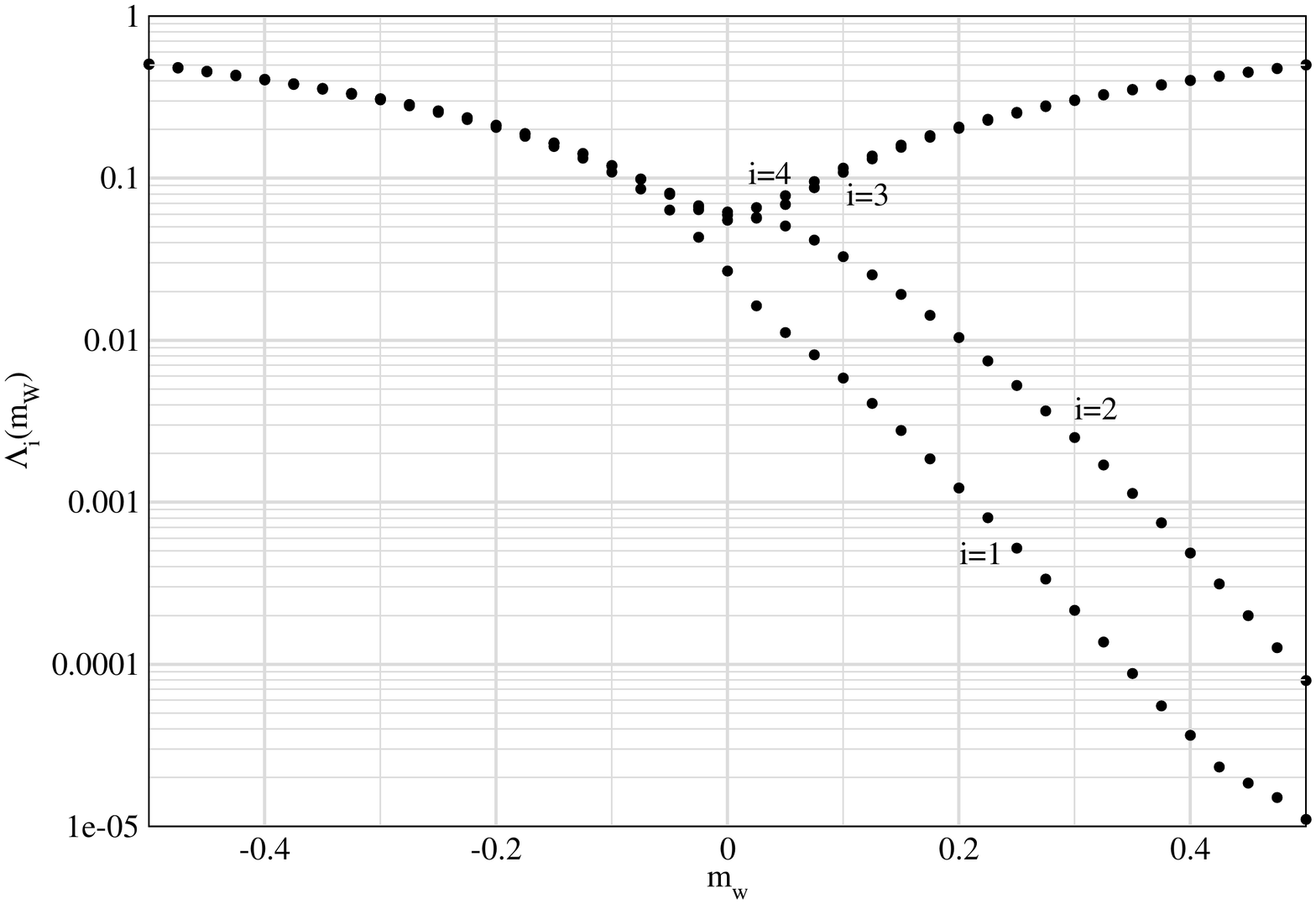}
\includegraphics[scale=0.275]{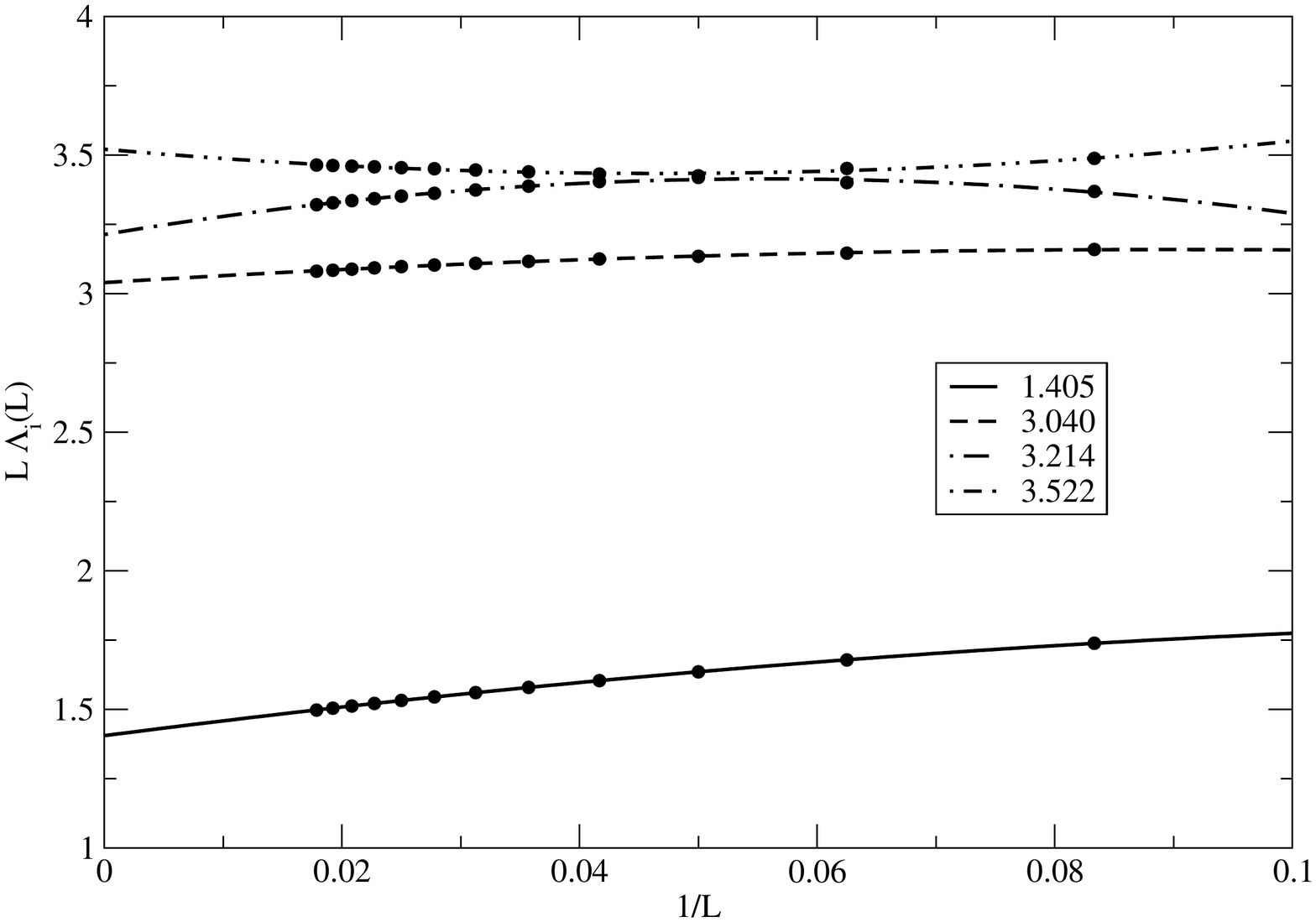}
\caption{The low lying eigenvalues, $\Lambda_i(m_w)$ as a function of $m_w$ at $L=56$
for a monopole-anti-monopole pair with $Q=2$ are shown in the left panel. The low lying spectrum at $m_w=0$ is shown as a function of $L$ in the right panel.
}
\eef{wilson-flow-q2}
In spite of the fact, that sensible results about monopoles could be obtained using overlap-Dirac fermions, we expect a numerical computation to
be difficult. The low lying eigenvalue(s) of the Wilson-Dirac operator that appears in the kernel of the overlap-Dirac operator will affect the
numerical computation. A study of compact QED using overlap-Dirac fermions is possible in principle but it will be numerically very expensive
to study such a theory due to the proliferation of low lying eigenvalues arising from a finite density of monopoles. 
This is evident in the left panel of \fgn{wilson-flow-q2} where the low lying eigenvalues of the Wilson-Dirac operator
as a function of Wilson-Dirac mass is plotted in the presence of a monopole-anti-monopole pair with $Q=2$. There are two anomalously small
eigenvalues for $m_w > 0$. In addition, the splitting of the two-fold degenerate spectrum is now seen in the lowest four eigenvalues of the massless Wilson-Dirac operator
as shown in the right panel of \fgn{wilson-flow-q2}. Therefore, both anomalous effects increase with $Q$. Yet, we expect the massless overlap-Dirac
operator to exhibit proper behavior as long as the numerical evaluation of the operator is performed accurately.

In spite of the anomalous behavior of the low lying eigenvalues of the Wilson-Dirac operator, the massless operator
produced the expected dimension of the monopole operator in~\cite{Karthik:2018rcg}. This is probably due to the fact that the entire
spectrum contributes to the dimension of the monopole operator and only the two lowest eigenvalues show a splitting of the two-fold degeneracy. Therefore, a cheaper alternative
would be to proceed in the same direction and compute the dimension of the monopole operator in non-compact QED
using Wilson-Dirac fermions in a fixed monopole-anti-monopole background and a computation in this direction is currently in progress.

\begin{acknowledgments}
  R.N. acknowledges partial support by the NSF under grant
number PHY-1515446.  N.K. acknowledges support by the U.S. Department
of Energy under contract No. DE-SC0012704.
\end{acknowledgments}

\bibliography{../../mynotes/biblio}
\end{document}